# DETECTION TECHNIQUES OF SELECTIVE FORWARDING ATTACKS IN WIRELESS SENSOR NETWORKS: A SURVEY

Preeti Sharma[1], Monika Saluja[2] and Krishan Kumar Saluja[3]

[1,2]Department of Computer Engineering, SBSCET Ferozpur, INDIA
engg.preeti@gmail.com
monika.sal@rediffmail.com
[3]Department of Computer Engineering, PIT Kapurthala, INDIA
k.saluja@rediffmail.com

## ABSTRACT

*The wireless sensor network has become a hot research area due its wide range of application in military and civilian domain, but as it uses wireless media for communication these are easily prone to security attacks. There are number of attacks on wireless sensor networks like black hole attack, sink hole attack ,Sybil attack ,selective forwarding attacks etc. in this paper we will concentrate on selective forwarding attacks In selective forwarding attacks, malicious nodes behave like normal nodes and selectively drop packets. The selection of dropping nodes may be random. Identifying such attacks is very difficult and sometimes impossible. In this paper we have listed up some detection techniques, which have been proposed by different researcher in recent years, there we also have tabular representation of qualitative analysis of detection techniques*

## KEYWORDS

*Wireless sensor Network, detection techniques*

## 1. INTRODUCTION

Wireless Sensor Network have become interesting and promising area of research and development ,we can define wireless sensor network as a self-configuring network of small sensor nodes which communicate with each other via radio signals and deployed in quantity to sense, monitor and understand the physical world .WSN combines sensing, computation and communication in a single device called sensor node. Wireless sensor nodes are also called motes. Sensor nodes have capability to collect sensed data and send that to the base station, a WSN generally consist of a base station that can communicate with a number of wireless sensors via radio link. WSN uses a wireless channel to communicate, so there are inevitably some issues such as message interception, tampering and other security [1]. Therefore, the security of networks has an important impact on the performance of monitoring, system availability, accuracy, and scalability, etc. The security of wireless sensor networks is an area that has been researched considerably over the past few years. The conventional security measures are not suitable to this wireless sensor networks due to resource constraints of both energy and memory. However, they are also highly susceptible to attacks, due to the open and distributed nature of the networks and the limited resources of the nodes. An adversary can compromise a sensor node, alter the integrity of the data, eavesdrop on messages, inject fake messages, and waste network resources. A common attack in WSN is DoS attack, and the objective of the attacker in DoS attack is to make target nodes inaccessible by [2] legitimate users. Many different kinds of DoS attacks against wireless sensor networks have been identified so far, e.g. selective forwarding attack, sinkhole attack, wormhole attack, black hole attack and hello flood attack, etc.

In this paper we will focus on selective forwarding attacks. In a selective forwarding attack[3], malicious nodes behaves like black hole and may refuse to forward certain messages and simply drop them, ensuring that they are not propagated any further. However, such an attacker runs the risks that neighbouring nodes will conclude that it has failed and decide to seek another route. A more subtle form of this attack is when an adversary selectively forwards packets. An adversary interested in suppressing or modifying packets originating from a few selected nodes can reliably forward the remaining traffic and limit suspicion of its wrongdoing. Selective forwarding attack can affect a large number of multi-hop routing protocols, such as TinyOS beaconing, directed diffusion [4], GPSR [5], GEAR, and can bring severe threats to the normal operation of the whole network, especially when it is used in combination with other attacks such as wormhole attack and sinkhole attack. This paper will give a review to various Selective Forwarding Attacks detection techniques. Also we have tabular representation of Qualitative Analysis of Detection Techniques.

## 2.PROPOSED DETECTION TECHNIQUES OF SELECTIVE FORWARDING ATTACKS

B.Yu [6] proposes a method to detect selective forwarding attacks based on checkpoints. Firstly choosing some nodes along the path randomly as the checkpoints node, then after receiving data packets, there will generate corresponding acknowledgments and then transmit them to the upper way. If any checkpoints node doesn't get enough acknowledgments, it will generate warning messages to the source node, so that the detection of the selective forwarding attacks can be realized. But an apparent problem exists in this process is that the nodes have to send acknowledgments continuously, which will greatly increase the cost of the network. By the way, this method can't judge whether there malicious tamper action exists.

Jiang [7] proposes a method to detect selective forwarding attacks, which is based on the level of trust and packet loss. After networking topology being established, when sensing data is transmitted on the path, the intermediate nodes detect and count the number of the packets they receive and send, and report the statistical results to the BS; According to these data, the BS calculates the trust level of nodes and evaluate the packet loss, so that it can determine whether this node is an active attacking node

Yu and Xiao in [8], proposed a scheme which uses a multi-hop acknowledgment scheme to launch alarms by obtaining responses from intermediate nodes. Each node in the forwarding path is incharge of detecting malicious nodes. If an intermediate node detects a node as malicious in its downstream/upstream, then it will send an alarm packet to the source node/base station through multi-hops

Sophia Kaplantzis et al [9] proposed a centralized intrusion detection scheme that uses only two features to detect selective forwarding and black hole based on Support Vector Machines (SVMs) and sliding windows. This intrusion detection is performed in the base station and hence the sensor nodes use no energy to support this added security feature. From this they conclude that the system can detect black hole attacks and selective forwarding attacks with high accuracy without depleting the nodes of their energy.

Brown and Xiaojiang [10] have proposed a scheme to detect selective forwarding using a Heterogeneous Sensor Network (HSN) model. The HSN consists of powerful high-end sensors (H-sensors) and large number of low-end sensors (L sensors). After deploying sensors, a cluster formation takes place with H-sensor as cluster head.

Xin, etal. Proposed [11] a light weight defense scheme against selective forwarding attack which uses neighbour nodes as monitor nodes. The neighbour nodes (monitoring nodes) monitor the transmission of packet drops and resend the dropped packets. They used a hexagonal WSN mesh topology.

Zurina Mohd Hanapi et al [12] proposed the dynamic window stateless routing protocol DWSIGF that is resilience to black hole and selective forwarding attack caused by the CTS rushing attack. Even without inserting any security mechanism inside the routing protocol, the dynamic window secured implicit geographic forwarding (DWSIGF) still promise a good defense against black hole attack with good network performance.

Riaz Ahmed Shaikh et al [13] proposed two new identity, route and location privacy algorithms and data privacy mechanism that addresses the challenging problem due to the constraints imposed by the sensor nodes, sensor networks and QoS issues. The proposed solutions provide additional trustworthiness and reliability at modest cost of memory and energy. Also, they proved that their proposed solutions provide protection against various privacy disclosure attacks, such as eavesdropping and hop by- hop trace back attacks.

Guorui Li et.al [14] has proposed the sequential mesh test based detection scheme. The cluster head node detects the packet drop nodes based on the sequential mesh test method after receiving the packet drop reports. This scheme extracts a small quantity of samples to run the test, instead of regulating the total times of test in advance. It decides whether continue the test or not based on the test result until it obtains the final conclusion. It requires less communication and computation power and shorter detection time to detect the selective forwarding attack nodes.

Deng-yin ZHANG et.al [15] et.al proposed a method to detect selective forwarding attacks based on digital watermarking technology. This method embeds watermark into the source data packets, which will be extracted at the base station (BS). The BS will judge whether there are malicious nodes in the transmission path by analyzing the packet loss rate from received data. Simulation results show that this method can effectively detect whether malicious nodes have discarded or tampered the contents of the packets.

**Table 1 Qualitative Analysis of Detection Techniques**

| Detection technique | Counter other attacks | Scheme nature | Consider other means of packet dropping | Multi path based | Acknowledgment Based | Neighbor monitoring | Reliable data delivery |
|---|---|---|---|---|---|---|---|
| Yu & xiao 's technique | No | Distributed | Yes | No | Yes | No | No |
| CHEMAS | No | Distributed | Yes | No | Yes | No | No |
| Support vector machine | Yes | Centralized | No | No | No | Yes | No |
| Krontirs et.al 's scheme | Yes | Distributed | No | No | No | Yes | |
| Two hop neighbor knowledge based scheme | No | Distributed | Yes | No | No | Yes | No |
| CADE | Yes | Centralized | No | No | Yes | No | No |
| Jremy brown et.al's technique | No | Centralized | Yes | No | No | Yes | No |
| Water mark based technique | No | Distributed | Yes | No | No | No | No |
| Chanatip et.al 's scheme | Yes | Centralized | No | No | No | Yes | No |
| Wang xingsheng et.al scheme | No | Distributed | No | No | No | Yes | Yes |
| Game theory model based scheme | No | Centralized | No | No | Yes | No | No |
| A sequential mesh test based scheme | No | Centralized | No | No | No | Yes | No |

## 3. CONCLUSIONS

Security and timely transmission of packets in wireless sensor network is its basic need of the network. The attack which affect this is the selective forwarding attack as in this attack malicious node drops the packet and make it unavailable to the destination. The detection of this type of attacks is important to meet the basic need of the network. Here is this paper we list up some detection techniques, which would help the user to know the techniques which have been proposed in recent year and in what way new techniques can be designed. This analysis will help us to know the previous proposed schemes and may helpful to design new one in the future.

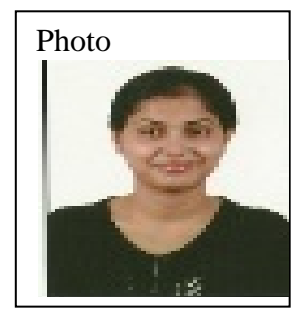

Short Biography- Preeti Sharma has done her B.Tech in computer science engineering from PTU currently she is doning her M.tech in computer science engg. from PTU. Her research area is wireless sensor networks.

Short Biography-

Monika Saluja has done BTech computer science and engineering from National Institute of Technology NIT, Jalandhar in 1997. She finished her MS software systems from BITS Pilani in 2002. Currently she is a PhD student in Department of Computer Science and Engineering a Guru Nanak Dev University, India.

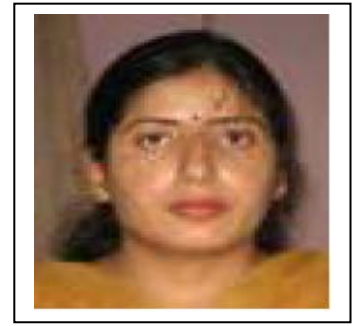

Short Biography-

Krishan Kumar has done BTech computer science and engineering from National Institute of Technology NIT, Hamirpur in 1995. He finished his MS software systems from BITS Pilani in 2001. Recently 2008, he finished his PhD from Department of Electronics and Computer Engineering at Indian Institute of Technology, Roorkee. Currently, he is an associate professor at PIT Kapurthala, India

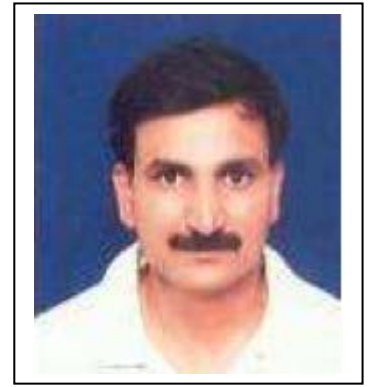